# Beyond Accuracy: Rethinking Hallucination and Regulatory Response in Generative AI

Zihao Li[1,2]; Weiwei Yi[1]; Jiahong Chen[3]

[1] CREATe Centre, School of Law, University of Glasgow
[2] Stanford Law School, Standford University
[3] School of Law, University of Sheffield

**Abstract**
Hallucination in generative AI is often treated as a technical failure to produce factually correct output. Yet this framing underrepresents the broader significance of hallucinated content in language models, which may appear fluent, persuasive, and contextually appropriate while conveying distortions that escape conventional accuracy checks. This paper critically examines how regulatory and evaluation frameworks have inherited a narrow view of hallucination, one that prioritises surface verifiability over deeper questions of meaning, influence, and impact. We propose a layered approach to understanding hallucination risks, encompassing epistemic instability, user misdirection, and social-scale effects. Drawing on interdisciplinary sources and examining instruments such as the EU AI Act and the GDPR, we show that current governance models struggle to address hallucination when it manifests as ambiguity, bias reinforcement, or normative convergence. Rather than improving factual precision alone, we argue for regulatory responses that account for language's generative nature, the asymmetries between system and user, and the shifting boundaries between information, persuasion, and harm.

# 1. Introduction

The rapid development and deployment of generative AI, particularly large language models (LLMs), have transformed the way information is generated, disseminated and consumed. As these models become increasingly integrated into critical domains such as healthcare, education, and law, their outputs carry significant epistemic, legal, and societal implications. One of the new risks posed by LLMs is hallucination, where the generated output is fabricated, nonsensical, subtle inaccurate, oversimplified, sycophancy or biased, yet delivered in a very confident tone (Ji et al. 2022; Li 2023; Wachter et al. 2024; Hacker et al. 2025a). Drawing on philosophy of science, such a phenomenon has been conceptualised as "bullshit" (Hicks et al. 2024). Hallucination has caused various types of practical harms, such as misinformation (Sun et al. 2024), disinformation (Bandara 2024), defamation (Binns and Edwards 2024), privacy infringement (Yao et al. 2024; Solove 2025) and even more serious mental and physical damages. Examples of the individualised and collective harms and damages abound, such as when a Norwegian user asked ChatGPT if it had any information about him, the chatbot made up a story that he had murdered his children (noyb 2025a). Similarly, it has been reported that ChatGPT falsely accused an American law professor by including him in a generated list of legal scholars who had sexually harassed someone, citing a non-existent The Washington Post report (Sankaran 2023). Even before the ChatGPT went viral, hallucination had already caused significant issues. A passenger booked a flight on Air Canada's official website, inquired about special discounts through a chatbot, which incorrectly told him he could apply for reimbursement after flying (Belanger 2024).

One of the common regulatory approaches, widely recognised among regulators, academia and technologists, is to prioritise *accuracy* as a primary metric for assessing the reliability of AI-generated content. For example, the UK ICO (the UK's data protection authority) consultation on its policy position on generative AI has a clear focus on accuracy of training data and model outputs, which emphasises how the accuracy principle applies to the outputs of generative AI models, and the impact that the accuracy of training data has on the output. The pursuit of accuracy to combat hallucination is further highlighted by the ICO's response (UK Information Commissioner's Office 2024; Li et al. 2024). Similarly, the European Data Protection Board's report concerning LLMs privacy risks reiterates accuracy as a common metrics to mitigate hallucination by which the factual errors could be reduced (Barberá 2025, p. 18). In the context of Agentic AI, the EDPB's report further subdivides the accuracy metrics into task-specific accuracy and step-level accuracy to gauge the coherence and correctness of intermediate steps and the final outcome. Similarly, the European Data Protection Supervisor's[1] guideline stresses the importance of ensuring the accuracy of structure and content of the training datasets, stating that "metrics on statistical accuracy (the ability of models to produce correct outputs or predictions based on the data they have been trained on), when available, can offer an indicator for the accuracy of the data the model uses as well as on the expected performance"(Barberá 2025, p. 15).

Many EU national Data Protection Authorities (DPAs) have also issued guidance in which accuracy is given paramount importance. For example, Belgium DPA's report explains that organisations must take reasonable steps to ensure the implementations of accuracy of personal data and require high-risk AI systems to use high-quality and unbiased data to prevent discriminatory outcomes and unintended behaviours (Data Protection Authority of Belgium 2024). Similarly, the Swedish DPA explicitly frames the GDPR's accuracy principle as both a normative standard and a practical regulatory tool to mitigate the risks of hallucinations in generative AI (Swedish Data Protection Authority (IMY) 2025). By highlighting accuracy-related measures (e.g., retrieval-augmented generation (RAG), confidence thresholds, and human review), the report positions accuracy not merely as a compliance obligation but as a dynamic mechanism for reducing the production of factually incorrect or misleading content (Swedish Data Protection Authority (IMY) 2025). Similar issue has also been raised by Non-Governmental Organization (NGO), such as noby, to fight against LLMs firms on the ground that their systems provide inaccurate information (noyb 2025b).

[1] The EDPB is an independent EU body established to ensure the consistent application of the GDPR. The EDPS is an independent supervisory authority responsible for monitoring the processing of personal data by EU institutions and bodies. The EDPS is a member of the EDPB.

A plethora of tech firms and third-party organisations similarly position accuracy as an unmitigated goal. As illustrated in Figure 1, OpenAI frequently highlights accuracy when announcing new models, using it as a proxy to indicate reductions in hallucination rates. Likewise, Google's Gemini platform presents accuracy as a marker of enhanced performance in its launch of Gemini 2.5 (Google 2025). Third-party organisations are no exception in the evaluation of LLM models. As shown in the Figure 2 below, accuracy is also employed as a key benchmark to assess and compare the hallucination rates across different LLMs.

Deeper world knowledge

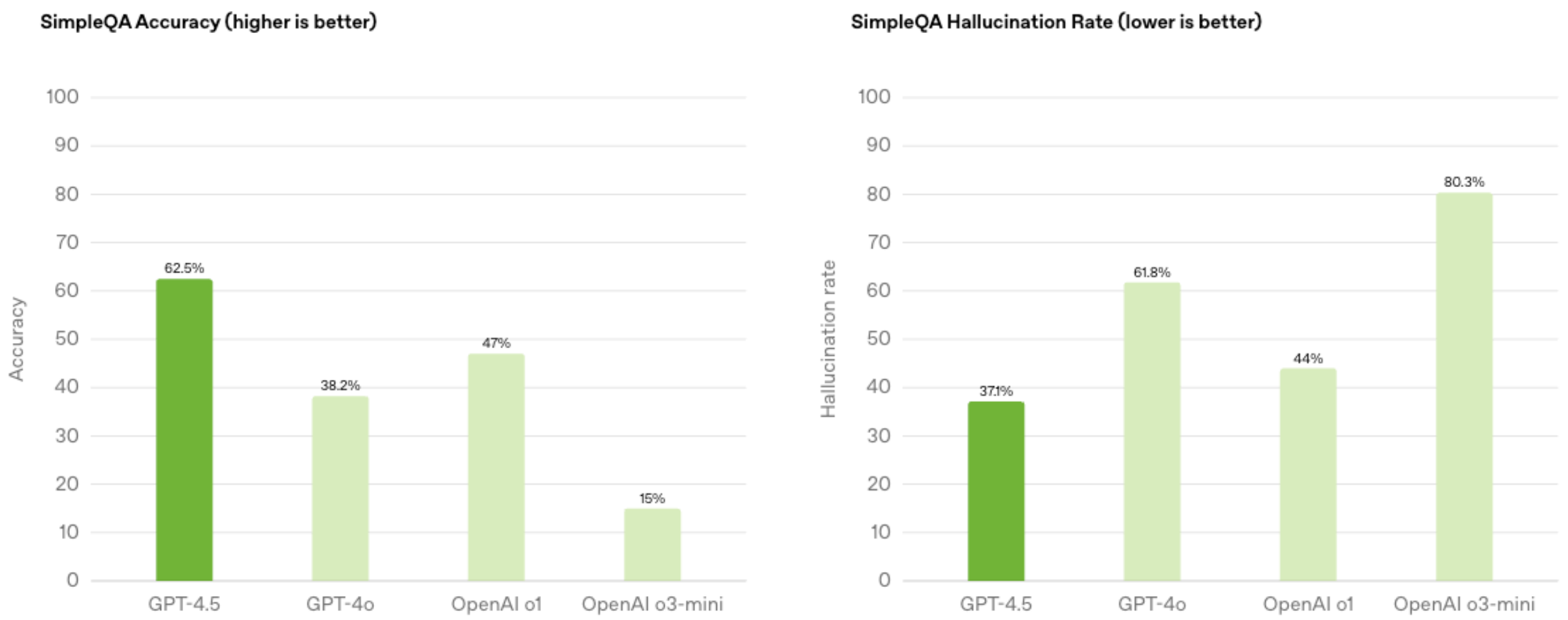

Figure 1 OpenAI Accuracy Demonstration (OpenAI 2025a)

Our benchmark revealed that Anthropic Claude 3.7 has the lowest hallucination rate (i.e. highest accuracy rate) of 17%.

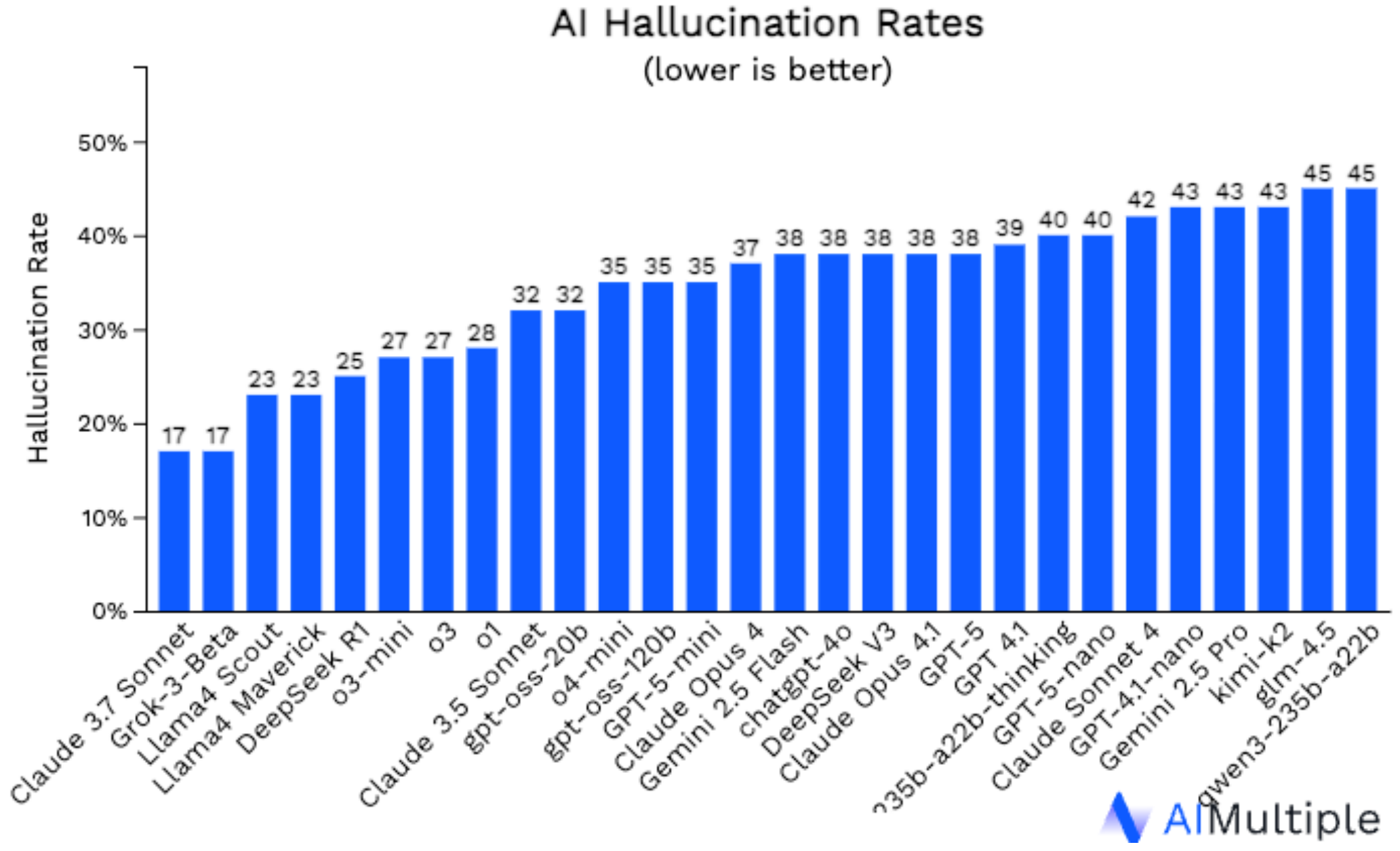

Figure 2 Third Party Assessment of LLMs (AI Multiple 2025)

As a result, accuracy has become an implicit assumption, either as an unchallenged good or as a technical necessity, based on the belief that high accuracy can mitigate hallucination and thereby ensure the responsible adoption of AI systems. Admittedly, improving accuracy can, to some extent, alleviate hallucination in LLMs, particularly by enhancing the factual reliability of model outputs, but the extent to which this should be translated into policymaking and regulatory objectives has not been sufficiently interrogated. In this regard, this article aims to conceptualise the ramifications of an accuracy-centred regulatory approach. We argue that an overreliance on accuracy as the primary benchmark for mitigating hallucination risks is both conceptually narrow and normatively insufficient. It risks obscuring a broader set of concerns, including the pursuit of epistemic truth, the trustworthiness and interpretability of information, and the diversity and heterogeneity of viewpoints. By framing hallucination solely as a technical flaw to be "fixed" through accuracy enhancement, we might risk overlooking the nuanced sociotechnical context under which information is constructed, received, and interpreted.

This paper advances the concept of *accuracy paradox* to describe such phenomenon. That is, the very efforts aimed at reducing hallucination through hyper-optimisation and overreliance of accuracy can exacerbate existing harms or give rise to new ones. In this paradox, the closer a system comes to only mimicking factual authority through enhanced accuracy, the more it risks creating a false sense of epistemic certainty, amplifying users' blind trust and weakening essential checks and balances. Thus, while accuracy may reduce hallucination in a narrow, statistical sense, it may simultaneously exacerbate deeper informational and cognitive vulnerabilities. This is not to suggest that improving accuracy is undesirable or the model developers should refrain from enhancing it. Rather, this calls for a more pluralistic approach to AI governance, one that complements accuracy with safeguards for epistemic trustworthiness, user interpretability and autonomy, and broader social good.

This paper first introduces the concept of hallucination by proposing a taxonomy situated within a socio-technical context, which also elucidates the underlying causes of hallucination. Secondly, the paper will conceptualise the accuracy paradox to explain how an overemphasis on accuracy, while seemingly beneficial, can obscure deeper epistemic, manipulative, and broad social concerns, ultimately undermining the responsible development and deployment of LLMs. Thirdly, the paper examines the existing EU regulatory frameworks in the context of accuracy paradox to evaluate how current legal instruments, such as the GDPR, DSA and AI Act, operationalise accuracy, and whether they adequately account for the accuracy paradox and its implications. Finally, based on conceptual analysis and policy implication, the paper will explore interdisciplinary solutions to mitigate the risks of LLM hallucinations that move beyond narrow accuracy metrics. These include technical strategies such as uncertainty expression, as well as legal and governance mechanisms that promote user empowerment, transparency, and polycentric oversight. Together, these approaches aim to foster a more trustworthy, epistemically reliable, and socially responsible LLMs system.

## 2. Hallucination: Concepts, Taxonomy, and Reasons

To put it simply (and perhaps overly-simplistically), the main logic behind the LLMs is to tokenise the text and data retrieved from training materials and to predict which string of words has the most likely sequence of words coming next (Li 2023; Binns and Edwards 2024). This probabilistic architecture, grounded in statistical correlations rather than grounded understanding, is the primary reason why hallucination emerges as an inherent feature of LLM outputs (Li 2023). Typically a model is further fine-turned via “reinforcement learning from human feedback” (RLHF) to ensure the consistency, coherence of model’s output and make their outputs more human-like and persuasive. However, such mechanism cannot guarantee the quality of generated content and does not incentivise the model to seek factual or trustworthy output, as it only predicts the likelihood of the next words, rather than comprehending, pondering and reasoning the underlying meaning or context behind the training data or users’ prompt (Zucchet et al. 2025). Moreover, RLHF could further exacerbate this effect as it imperceptibly introduces latent annotator subjective feedback and reinforces the model’s awarding mechanism, which may unintentionally amplify inaccurate, biased (Bai et al. 2025) or misleading responses. RLHF has been criticised for having shaped AI into a sycophantic, people-pleasing persona that mirrors the human reluctance to confront uncomfortable truths: When AI constantly apologises, avoids offense, and earnestly fabricates answers to preserve our feelings, it lays down a series of gentle traps: subtle, well-intentioned, but ultimately deceptive (Sharma et al. 2024). Such behaviour exemplifies sycophancy hallucination, where outputs are not directed towards epistemic reliability but towards maximising user comfort and satisfaction, producing responses that appear reassuring yet remain epistemically hollow and untrustworthy.

Especially in zero-shot contexts, where the LLMs do not have a specific training dataset for the given task or domain, the model is forced to rely on general patterns learned during pretraining, which can increase the likelihood of hallucinations, inaccuracies, or contextually inappropriate responses (Kojima et al. 2023). Meanwhile, any misinterpreting prompts may further exacerbate the risks of hallucination. Without a proper and specific prompt to guide models in zero-shot setting, the inherent ambiguity may lead the model to overconfidently generate plausible-sounding yet unfounded outputs, thereby compounding the epistemic instability of its responses.

LLMs not only fabricate or generate “nonsense” content, but also prone to produce subtle mistruths (sometime combined with real facts), alongside oversimplifications of complex topics and responses biased towards certain commonly held beliefs (Bender et al. 2021; Li 2023; Wachter et al. 2024). It has also been documented in literature that the way that LLMs express and interact with users are hardly reflection of their internal workings, as they tend to convey information in a very confident way, which could misguide professional experts, let alone lay people (Chen et al. 2025).

It should be noted that hallucinations can take different forms, and there is a need to differentiate various types of hallucination. Building on the work of Huang et al. (2025), and drawing on the wider literature, we present a taxonomy of hallucination, as outlined in Table 1.

| Category | Type | Descriptions |
|---|---|---|
| Factuality Hallucination | Factual Contradiction | Factual errors occur when a model generates information that is objectively incorrect or inconsistent with established facts. It includes |
| | | (1) Entity-error hallucination: erroneous entities, e.g., wrongly saying Thomas Edison invented the telephone;<br>(2) Relation-error hallucination: wrong relations, e.g., misstating Edison's role in inventing the light bulb. |
| | Factual Fabrication | Factual fabrication refers to the generation of entirely false or invented information by a model, which does not correspond to any real-world facts or data.<br>(1) Unverifiability hallucination: Unverifiable statements, e.g., Eiffel Tower construction led to Parisian tiger extinction (non-existent species);<br>(2) Overclaim hallucination: Claims lacking universal validity, e.g., Eiffel Tower sparked global green architecture movement. |
| | Conflation of Facts | Conflation of facts occurs when a model combines or merges distinct pieces of information, creating a distorted or misleading representation of reality. This often involves blending facts from different contexts, events, or sources in a way that makes them appear as if they are connected or part of a single narrative, even when they are not. |
| Consistency Hallucination | Instruction Inconsistency | Outputs deviating from user directives, e.g., answering instead of translating. |
| | Context Inconsistency | The output is inconsistent with the provided context, such as adding information that is not present in the given context. |
| | Logical Inconsistency | Internal contradictions in reasoning, e.g., 9.11>9.8 |
| Reference Hallucination | Source Fabrication | This occurs when the model references or cites sources that do not actually exist or are fabricated |
| | Misattribution | This happens when the model attributes information or quotes to the wrong source, either inaccurately referencing the original source or incorrectly attributing content to someone or something else. |
| Sycophancy Hallucination | Overly Complimentary | This type occurs when the model generates content that excessively praises a person, idea, or organization, often beyond what is warranted or realistic, in order to align with perceived preferences or authority figures. |
| | Flattery Bias | Model produces outputs that are skewed to present the person or entity in the most favourable light, possibly distorting facts or providing biased interpretations to maintain favour or avoid conflict. |
| | Subservient Tone | The model generates content that adopts a deferential or excessively accommodating tone, suggesting compliance or agreement with a particular individual or viewpoint in an exaggerated way. |
| Consensus Illusion | Narrowed Perspective | The generated output only reflects a narrow or limited perspective, giving the false impression that there is a broad consensus or agreement on an issue. However, there may be significant disagreement or diversity of opinion. |
| Oversimplified Hallucination | The generated content that reduces complex or nuanced information to overly simplistic terms, ignoring important details, context, or variations. This often results in misleading or incomplete answers that may appear clear and straightforward but fail to accurately reflect the complexity of the issue | |
| Prompt-Sensitivity Hallucination | Sandbagging | LLM automatically reduces the quality of its responses in reaction to prompts perceived as "lower quality", such as those with poor grammar, typos, or informal structure. |
| | Emotionally induced drift | A tendency for LLMs to produce misaligned or overly reassuring outputs in response to emotionally charged prompts, regardless of the factual or epistemic grounding of the response. |

Table 1 Taxonomy of hallucinations

From Table 1, it is clear that the sociotechnical complexity of hallucination extends far beyond mere factual inaccuracies. It involves inconsistent responses, reference fabrication, and more subtle psychological and social issues such as sycophancy tendency and consensus illusion. The diversity of AI hallucination means that a narrowly defined concept of accuracy may not be fit for purpose when guiding regulatory responses. In the next section, we provide a theoretical account on why this might be the case by introducing the concept of accuracy paradox.

# 3. Understanding the Accuracy Paradox

Over-reliance on improving accuracy of LLMs does not always reduce harm and may, in certain contexts, exacerbate it. While accuracy is conventionally treated as a normative and unquestioned good, this section interrogates its limits and unintended consequences. We use the term *accuracy paradox* to describe the scenario where pursuing higher accuracy of LLMs by model developers and policymakers with a view to addressing hallucination harms, with a narrow understanding of accuracy, may paradoxically oversee or even give rise to more subtle forms of harms. We demonstrate this phenomenon by contrasting accuracy with three dimensions of desirable policy goals of AI regulation: *output trustworthiness*, *individual autonomy* and *social progression*.

## 3.1. Accuracy vs. Trustworthiness

The notion of trustworthiness, often invoked as an implicit normative aim of AI regulation, contains a compound epistemic and procedural value (Laux et al. 2023).[2] In the context of LLMs, this article differentiates three analytically distinct but interrelated dimensions of trustworthiness: epistemic validity, user perception, and interpretability. First, trustworthiness entails that an output must not only be statistically accurate but also epistemically grounded, anchored in justified reasoning, verifiable knowledge or even falsifiable argument (Section 3.1.1). In philosophical terms, this links trustworthiness to the pursuit of truth, not as mere factual correspondence, but as a normative value requiring justification and resilience to error, thereby distinguishing reliable knowledge from outputs that only appear accurate (Popper 1959; Williams 2002). Second, trustworthiness concerns user psychology and behavioural reliance: outputs that appear linguistically accurate may induce unwarranted trust, particularly when users forego critical scrutiny (Section 3.1.2). Third, trustworthiness depends on transparency and interpretability. Users must be able to trace the reasoning process or access uncertainty signals in order to make informed judgments about the credibility of a given response (Section 3.1.3). Taken together, these three perspectives demonstrate that trustworthiness is not reducible to accuracy alone. Rather, trustworthiness emerges from the interplay between internal epistemic mechanism, output quality, user trust calibration, and the visibility of the epistemic process that generated the output. Without such foundation, the regulatory promise of "trustworthy AI" risks collapsing into a rhetorical ideal rather than a meaningful standard. The following subsections examine each of these dimensions in turn.

### 3.1.1. Accuracy Is Not Truth: Epistemic Limits of Predictive LLMs

Higher accuracy requirements do not necessarily diminish the potential harms users receive, but may instead increase the risk of the users trusting misinformation. The primary reason for such a paradox is that accuracy is not equal to truth. From a legal and philosophical perspective, accuracy often refers to the consistency of information with a predetermined, up-to-date and "ground truth" dataset,[3] whereas truth is a more complex philosophical concept that involves a deeper, context-driven evaluation that aligns with both factual correctness, epistemic and social coherence. From a philosophical viewpoint, truth encompasses various schools of thought in philosophy, such as positivism, constructivism, correspondence theory, coherence theory and consensus theory (Dowden and Swartz 1995; Junior 2023). Over-reliance on accuracy oversimplifies the requirement of truth and trustworthy information, which may create epistemic overreach. Truth, in this broader philosophical sense, is not a singular or purely empirical concept but a contested and multi-dimensional construct. At its core, truth concerns the relationship between statements and reality, yet how that relationship is defined varies across traditions. Correspondence theory defines truth as the alignment between a statement and an objective fact or state of affairs (Dowden and Swartz 1995). Coherence theory considers a belief true if it fits consistently within a broader system of beliefs (Dowden and Swartz 1995). Consensus theory views truth as that which is agreed upon by a community under ideal conditions (Habermas 2005), while constructivism understands truth as shaped by social, cultural, or historical contexts (Kukla 2000, pp. 7–18). In contrast to these definitions, positivism grounds truth in empirical verification and logical consistency. While accuracy based on ground truth may achieve what is close to positivist truth, it is unlikely to fulfil the test in other theories as they concern not merely isolated correctness or accuracy but about justification, context, and interpretive frameworks—dimensions that statistical accuracy alone cannot adequately capture.

[2] Art. 1, EU Artificial Intelligence Act: https://eur-lex.europa.eu/eli/reg/2024/1689/oj/eng

[3] Article 4, EU General Data Protection Regulation (GDPR): https://eur-lex.europa.eu/eli/reg/2016/679/oj/eng

The pursuit of truth for LLMs as an ultimate goal is essential. Otherwise, LLMs may combine accurate fragments into a misleading whole due to a lack of contextual understanding or epistemic grounding, leading to factuality and consistency hallucinations. A technically "accurate" response can still mislead, misrepresent, or obscure important nuances, thereby creating an illusion of trustworthiness where none exists. In this light, accuracy without truth not only fails to resolve the problem of hallucination but may, paradoxically, deepen its epistemic and normative harms.

As illustrated in Section 2 on the mechanisms of LLMs, accuracy is judged by token prediction success or benchmark alignment, which is statistical, not epistemic. Therefore, the increase of accuracy in LLMs only enhances the syntactic or probabilistic plausibility of output, rather than their epistemic validity. In other words, a model could become better at predicting what sounds correct based on training data, without any grounding in whether the content is actually true, justified, or verifiable. This distinction is critical: a linguistically fluent and statistically probable output may still be epistemically hollow, particularly when the model lacks access to mechanisms for truth verification or source attribution.

It is perceivable that even when an LLM is based on a comprehensive ground truth corpus, its underlying generative architecture may still remain susceptible to hallucination. This is not merely a question of data coverage, but of architectural constraints. Fundamentally, given the probabilistic nature of token prediction, the model's output is a function of likelihood estimation rather than semantic understanding or truth retrieval. Errors can still emerge from token-level misalignment, prompt sensitivity, and the diffusion of meaning across latent spaces. In other words, even if every "correct" answer exists somewhere in the model's training data, there is no guarantee that the model will retrieve, compose, or represent it truthfully in response to any given prompt. This reveals a fundamental fragility: the hallucination problem cannot be resolved solely by improving training data accuracy, because the architecture itself does not guarantee faithful grounding or truth-seeking. Over time, such outputs risk accumulating what Zittrain (2022) refers to as "intellectual debt": responses that appear authoritative yet obscure their own opacity, thereby undermining long-term epistemic trust. This explains why factuality, inconsistency and reference hallucination often occur.

Therefore, overreliance on accuracy amounts to an oversimplication of the truth requirement. Equating truth with accuracy as measured against the training data and output's consistency is inherently flawed. This approach fails to account for the depth and contextual nuance required to genuinely reflect truth, as it reduces the concept to a mechanical alignment with surface-level vectorised data without addressing broader epistemic values like coherence, social context, or representational fairness (See Section 3.2 and 3.3). It overlooks the complexity of truth as something that extends beyond statistical conformity to include deeper, context-driven trustworthiness, validation and reliability. As a result, merely improving accuracy masks the absence of truth, ultimately undermining the trustworthiness of LLMs' outputs.

### 3.1.2. Overreliance on Accuracy Can Create Over-trust

This misalignment of accuracy and truth leads to information that is often not trustworthy. High accuracy requirements can heighten the users' over-trust without critically evaluating it (Klingbeil et al. 2024, p. 2). As measured accuracy rises, users generalize from aggregate performance to instance-level reliability, accept outputs at face value, and verify less, which leads to new harms: over-trust of AI. In other words, merely improving the accuracy of the models is insufficient, because the more accurate the model is, the more users will rely on it, and thus be tempted not to verify the answers, leading to greater risk when hallucinations appear (Li 2023). This problem stems from the fact that LLMs are fundamentally designed to produce fluent and convincing responses, rather than inherently comprehending content, reasoning and pursuing truth as an epistemic goal. Even if a model achieves high accuracy, it cannot guarantee a fully trustworthy response because LLMs merely predict the likelihood of word sequences without truly understanding the content they generate. If even 0.1% or 0.2% of answers in a particular domain are incorrect, it creates significant challenges for users attempting to discern the authenticity of those responses, especially in legal, medical or other high-risk domains (Li 2023). Therefore, overreliance on accuracy could lead to a false sense of trustworthiness, where users assume outputs are fully reliable despite the presence of subtle but consequential errors.

In this regard, demonstrating accuracy directly to users could also mislead users and cause them to over-trust. A sole emphasis on accuracy rates foregrounds surface-level factual correctness, targeting only factual hallucinations, while obscuring deeper epistemic concerns. As highlighted above, the deployment of Reinforcement Learning through Human Feedback (RLHF) could introduce several concrete epistemic limitations. These limitations cannot be resolved through accuracy rate improvements alone; in fact, the pursuit of higher accuracy may further entrench the misleading nature of model outputs by masking their underlying uncertainties and reinforcing rhetorically plausible but potentially untrustworthy responses. RLHF embeds human labellers' subjective views, values and backgrounds into the system, which do not always reflect trustworthiness comprehensively. As a result, what may be perceived as "true" by these systems is often merely an outcome of probabilistic modelling and human reinforcement, rather than an inherent understanding or intent to pursue the truth. Generated responses, therefore, are more of a product of chance and reinforcement than any genuine alignment with truth. Worse still, fine-tuning through human feedback further exacerbates such flaws, as LLMs would inadvertently prioritise persuasive or rhetorically appealing responses over those that are truthful (Zhou et al. 2024). It often generates outputs that sound confident or convincing, rather than those grounded in verifiable and critical information (See Section 3.2). This creates an accuracy paradox: while efforts to enhance accuracy may improve the articulation and persuasiveness of outputs, they do not necessarily make them more trustworthy, likely reinforcing biases and creating the illusion of reliability without a genuine commitment to truth. In this context, improvements in accuracy can lead to over-trust and mislead users into abandoning doubt and independent judgment, creating a false sense of reliability that discourages scrutiny of the model's outputs (See Section 3.3).

#### 3.1.3. Accuracy-Transparency Trade-Off: Accuracy Can Undermine Meaningful Transparency and Interpretability

Overreliance on accuracy also often comes at the cost of transparency and interpretability of the generated output (Kossow et al. 2021). At first glance, enhancing accuracy appears to be a straightforward improvement in model performance. Yet, as models grow in complexity and parameter, the opacity of their internal reasoning deepens and it becomes more difficult to trace back the epistemic opacity (Chen et al. 2025), which undermines the transparency and interpretability of generated content. As exemplified by the OpenAI reasoning model o1, the chain of thought (CoT) and reasoning trace are often cloaked. Some users have reported getting warnings or even having their accounts blocked whenever their prompts include terms like "reasoning trace" or "show your chain of thought" (Awayyy 2024; Mindfultime 2024). It has been reported that when prompted to explain its reasoning, Claude fabricates plausible-sounding explanations and tailors its false reasoning to agree with misleading prompts (The Economist 2025). This further illustrates the "illusion of thinking": LLMs may pretend reasoning in ways that convince users of genuine thought, yet as Apple's recent work demonstrates, such reasoning often amounts to pattern matching rather than transparent and verifiable logical reasoning (Shojaee et al., pp. 5–7). It also demonstrates that CoT in Reasoning Models (LRMs), such as Claude 3.7 and DeepSeek-R1, fail to develop generalisable problem-solving capabilities and often collapse entirely when faced with higher-complexity tasks (Shojaee et al.). Even when provided with explicit algorithms to execute, these models fail to perform accurate step-by-step reasoning, revealing limitations in symbolic manipulation. Moreover, their "thinking traces" often include internally inconsistent or redundant reasoning, and models may even reduce their reasoning effort as problems become more complex. This illustrates the opacity of current LLMs with respect to revealing their internal decision-making processes, which in turn compromises users' ability to verify, question, or contest the system's outputs.

Assumptions that accuracy is an unmitigated good are contested in this context, as they overlook the complex trade-offs between surface-level correctness and deeper epistemic values such as transparency, interpretability, justification, and trustworthiness. Trustworthiness, in this regard, is not an inherent property of output, but an emergent requirement that users need to access and interpret the underlying rationale of the system's response. This makes meaningful transparency and interpretability critical preconditions: without knowing how or why a model arrived at a conclusion, users are deprived of the epistemic scaffolding necessary to evaluate its legitimacy or challenge its implications. Transparency alone does not guarantee this, as it is normally limited to disclosure of system components. Interpretability helps to complement this gap by enabling the tracing of reasoning, assessment of confidence, and identification of omissions, which renders outputs assessable, justifiable, and ultimately trustworthy. As a result, in such contexts, only improvements in accuracy may paradoxically erode trustworthiness, precisely because they induce users to accept outputs without the capacity to interrogate how those outputs were generated. When the reasoning process is inaccessible or obscured, users are deprived of the epistemic tools needed to evaluate the credibility of responses, encouraging passive acceptance rather than critical engagement. This mechanism undermines the normative foundations of trust, which relies not merely on outcome quality but on the ability to assess the integrity of the process by which that outcome is produced.

In the same vein, as discussed in Section 3.1.2, if accuracy improvements primarily manifest in enhanced linguistic fluency and rhetorical confidence, such convincingly expressed outputs may obscure underlying model uncertainty, particularly in zero-shot contexts where the model lacks task-specific training data. In these cases, the model's internal epistemic uncertainty is not meaningfully communicated through its surface-level expression (Chen et al. 2025). That is, even when the model has low confidence in its response, it often presents the output with syntactic precision and stylistic confidence, thereby concealing the absence of grounding or justification. This misalignment between internal uncertainty and external presentation undermines meaningful transparency and interpretability. Users are not given access to indicators of confidence or error margins, nor are they able to trace the model's reasoning in a way that would allow them to assess reliability (Yin 2025). In effect, the pursuit of accuracy, when limited to surface-level plausibility, reinforces the accuracy paradox: it renders the output more persuasive, thereby undermining transparency while simultaneously depriving users of the epistemic tools necessary for critical scrutiny, deliberation, and informed decision-making. As a result, in such contexts, improvements in accuracy may paradoxically diminish trustworthiness, as they mislead users by obscuring the reasoning process and preventing meaningful evaluation of the output's credibility.

### 3.2. Accuracy vs. Autonomy

Beyond output-level risks, when a user is exposed to an accumulation of responses over time, the hyper-focus on the statistical accuracy of LLM-generated outputs may also risk obscuring deeper concerns around manipulation, ultimately undermining users' autonomy. Three key dynamics drive this risk. First, statistical accuracy can function as a disguise. Regulatory metrics that emphasise accuracy incentivise developers to optimise surface-level fluency and rhetorical persuasiveness, rather than ensuring epistemic reliability. Users are thus not persuaded by sound reasoning, but by linguistically confident outputs, leading to a misplaced sense of trust in the model's reliability. This is linked to the trustworthiness issue flagged up in the previous section but pertains to a different concern: even if a series of outputs individually fulfil the "truth" criterion (however it is defined), they are not necessarily "neutral" in the sense that they do not steer the users to a particular behavioural or perceptual direction. Second, accuracy as a singular benchmark is inherently limited. It fails to capture the epistemic and contextual complexity of LLM outputs during dynamic interactions, particularly when responses are opinion-based, value-laden, or socially contingent. In other words, it cannot account for outputs that fall into the grey area of "not being inaccurate" responses that evade absolute factual error while still exerting potentially manipulative influence. Third, accuracy, as a static benchmark, fails to account for the dynamic deterioration that emerge through continued user interaction in evolving real-world contexts. Such static evaluation is ill-equipped to capture the effects of training-test contamination, which artificially inflates accuracy by rewarding memorisation, and the evolving dynamics of user interaction (e.g., prompt poisoning), which causes consistency and sycophancy hallucinations. When this fragility is mistaken for stable performance, it undermines users' ability for informed judgments and leaves them vulnerable to subtle manipulation and erode autonomy.

As a result, the more we rely on and optimise for accuracy, the more we risk overlooking the subtle ways in which persuasive yet opaque outputs can influence, nudge, or manipulate user behaviours, often without users being fully aware of such influence. The rest of this section will elaborate on each of the three kinds of risks in turn.

### 3.2.1. Accuracy vs. Epistemic Independence: The Rhetorical Illusion

When accuracy becomes a regulatory anchor, optimisation often targets linguistic fluency, rhetorical confidence, and audience fit as a means of influencing users, rather than enhancing models' epistemic reliability, internal understanding or reasoning capacity. This is because, as explained above, the most straightforward and computationally efficient way to signal reliability is to enhance linguistic fluency and persuasive tone, as these features that lead users to perceive the output as accurate and trustworthy. In such cases, the appearance of accuracy functions less as a marker of epistemic validity and more as a rhetorical disguise, one that can mislead users into overestimating the reliability of the content. As demonstrated by Okoso et al. (2025), AI expressions significantly change users' decisions. When interacting with AI-generated responses, users' choices were shaped by the tone of the output regardless of their prior knowledge or individual attributes. Older users were especially susceptible to tonal influence, and highly extroverted individuals often made decisions that diverged from their stated perceptions. Similarly, Salvi et al. (2025) demonstrate that when individuals lack strong prior opinions on a given topic, GPT-4 proves significantly more persuasive than human counterparts under conditions of microtargeting. Specifically, GPT-4 made participants 81.2% more likely to change their views towards its assigned position after the engagement, outperforming human persuaders in 64.4% of cases, Crucially, such rhetorical strength is not grounded in epistemic superiority or factual depth, but in the optimisation of surface-level linguistic features, further illustrating how accuracy enhancements may amplify user persuasion while obscuring the absence of transparent reasoning or verifiable information. Paradoxically, these surface-level improvements may erode users' epistemic independence by encouraging passive acceptance rather than critical engagement, thereby undermining individual's autonomy.

Such persuasive power of LLMs, when embedded within interactive systems, gradually shifts from persuasion to what Luciano (2024) refers to as hypersuasion: a technologically mediated mode of influence that empowers the system to identify, target, and exploit users' cognitive and emotional susceptibilities with unprecedented subtlety and reach. As suggested by Yeung (2017), such rhetorical strength gives rise to what is known as "hypernudge" where algorithmic architectures, continuously updated and highly personalised, can restructure the user's informational environment in real time to shape behaviour without the user ever realising it. Together, they point to a disturbing possibility: that AI systems, under the guise of fluency and relevance, may intentionally or unintentionally manipulate users beneath the threshold of awareness, not by being inaccurate, but precisely by being rhetorically too accurate.

### 3.2.2. Accuracy vs. Manipulation Resilience: The "Not Inaccurate" Blind Spot

Overstressing accuracy as a singular benchmark is inherently flawed, as it creates a false sense of trustworthiness that blinds users and regulators to the subtler manipulative risks embedded in outputs that are not being inaccurate. Unlike verifiable questions in mathematics, coding, or scientific facts (OpenAI 2024a), many answers generated by LLMs are opinion-based and lack standard or verifiable solutions, as even humans may hold differing views. These differences can arise from individual experiences, cultural backgrounds or beliefs, language use, and other contextual factors (Hacker et al. 2025b, p. 54). These responses, while may be free from overt factual errors, can still exert significant influence on user beliefs, behaviours, and decisions through rhetorical framing, selective emphasis, or tonal persuasion. This echoes to the fact that, as shown above in Table 1, there are multiple types of hallucinations beyond factuality hallucination, including consensus illusion, oversimplification, and consistency hallucinations.

These hallucinations emerge precisely because the model's responses tend to oversimplify facts and viewpoints, often producing small mistruths, reductive representations of complex issues, and outputs biased towards dominant narratives or widely held assumptions (Wachter et al. 2024, p. 2). Rather than reflecting epistemic rigor, such responses mirror statistical patterns in training data, thereby reinforcing prevailing perspectives while marginalising nuance and dissent. Unlike outputs that are demonstrably false, such consensus illusion, oversimplification, and consistency hallucinations caused by "not being inaccurate" answers are difficult to contest because they remain formally consistent with facts or norms. This also explains why over-relying on accuracy as a benchmark makes it particularly difficult to address these hallucinations. However, they can still shape perceptions through mechanisms such as downplaying opposing views, emotional tone manipulation, or subtly biased ordering of information. As Wang (2024) argues, LLMs generate coherence not from communicative intent, but from token-level prediction trained to maximise fluency. Users, nonetheless, tend to interpret this fluency as indicative of understanding or reliability. As a result, accuracy paradox occurs: improvements only in accuracy misplace user trust, as narrowly defined accuracy metrics fail to capture manipulative but technically "not inaccurate" outputs. This diminishes users' ability to critically assess content, thereby lowering autonomy and manipulation resilience. These risks are not merely theoretical. To explain this in more detail, we distinguish two different scenarios.

a. *Subtle manipulation by design: Ads powered by AI*

Such subtle manipulation can be by design, that is, the designers' intent of imposing particular hidden influences upon end-users by producing outputs that are "not being inaccurate", yet difficult to falsify. Advertisements embedded within seemingly neutral responses exemplify how LLMs may guide users towards specific products or services under the guise of helpfulness, constituting "not being inaccurate" content with commercial intent. These LLM-powered advertising has emerged in practice. OpenAI has announced to experiment with integrating advertising into ChatGPT, including product details, pricing, reviews, and even direct purchase links, as illustrated in Figure 3 (OpenAI 2025b). However, users have reported that ChatGPT nudged them towards purchasing a nutrition programme with a functional link during a conservation about sushi (Latestly 2025). This is not an isolated case. Other users have similarly claimed that ChatGPT recommended a cuisine-related ads during the discussion about traveling. While an OpenAI engineer has attributed these instances to hallucinations (Eleti 2025), it might be technically difficult to rule out the potential intention of the developers to influence users with the commercial information that are "not inaccurate" but subtly manipulative. The potential of such practice is mirrored by a recent collaboration by Amazon and Anthropic, who are developing more persuasive virtual assistants powered by Claude AI (Bensinger 2024), even though multiple deceptive design patterns had already been previously identified during user inactions with those virtual assistants (Owens et al. 2022). Recent study also shows that such subtle manipulation has influence on users' ideology (Paschalides et al. 2025).

Regardless of whether such responses are classified as hallucinations, they underscore the inadequacy of relying solely on accuracy as the benchmark for addressing LLMs' hallucinations. On the one hand, these illustrate that even outputs containing verifiable and accurate information, such as product names, pricing, and links, can still mislead users through subtle contextual misalignment or rhetorical nudging. On the other hand, they reveal how "not being inaccurate" can serve as a vehicle for embedded influence, where suggestive outputs appear plausible yet erode user autonomy and critical discernment. This highlights the necessity of incorporating broader evaluative metrics, such as contextual relevance, intent, and epistemic transparency beyond surface-level factual accuracy.

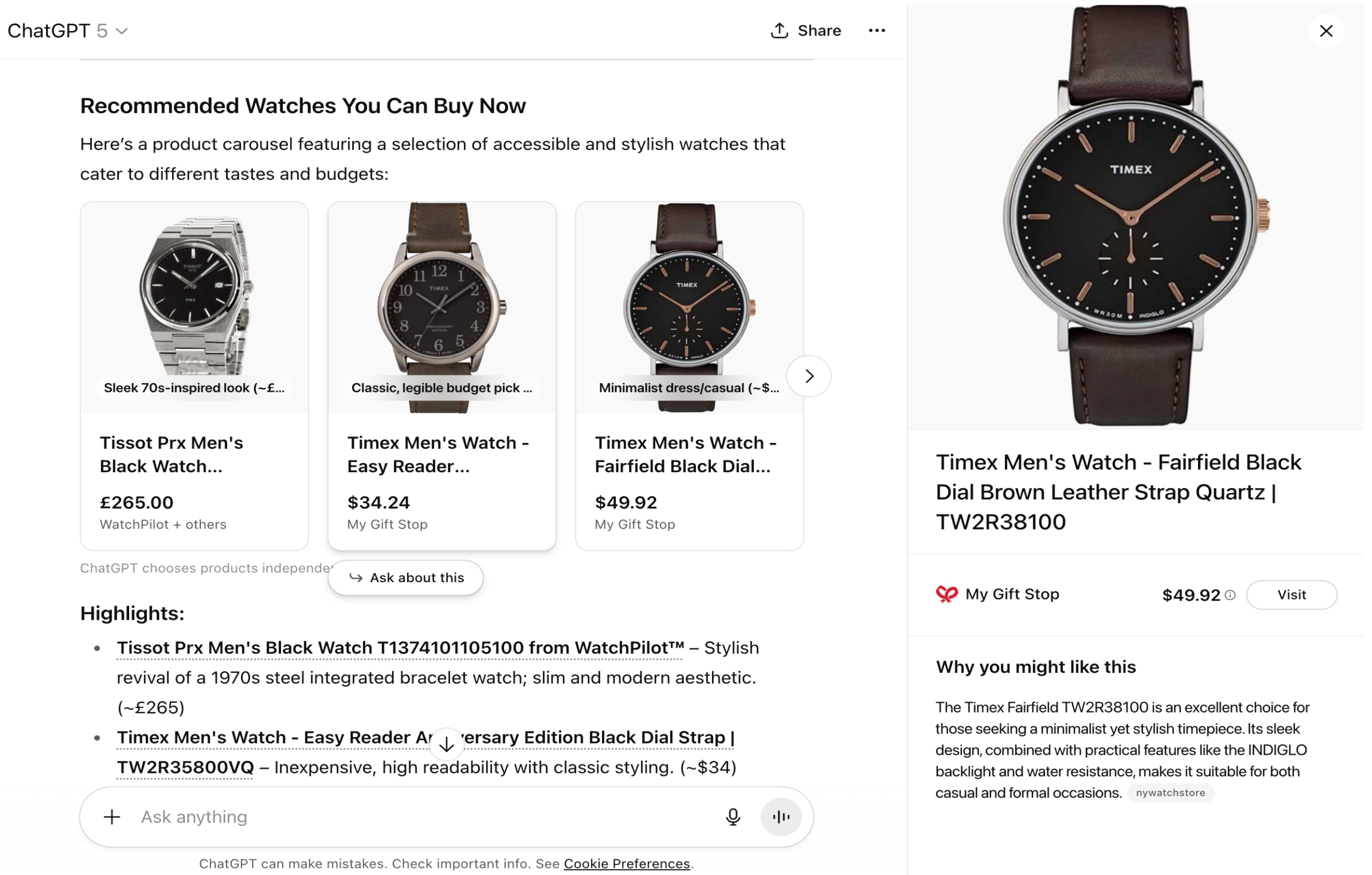

Figure 3: Screenshot of ChatGPT 5's commodities recommendations

*b. Personalised manipulation through the over-reliance of high accuracy*

At a more discreet level, overreliance on high accuracy may lead to user over-trust, and thus manipulation in a personalised manner. As discussed in Section 3.1.2, persuasive outputs coupled with statistical credibility encourage users to accept and trust information without interrogation. In contexts where users share personal data or engage repeatedly with a system, this can enable a more targeted and personalised manipulation, especially when interaction histories are long enough to allow tailored behavioural nudges based on linguistic patterns or psychological triggers. An LLM need not "intend" to manipulate; the manipulation emerges systemically from design choices, reinforcement learning, and stochastic modelling.

Collectively speaking, manipulation through LLMs introduces a new risk: the reciprocal exchange enables adjustment based on user input in real-time, while the long-context capacity allows more subtle influences through unnoticeable increments, while the vast quantity of interaction may put extra challenges to human reviewer in detecting the manipulation (El-Sayed et al. 2024, p. 5).

### 3.2.3. Static accuracy vs. Dynamic Interaction: Ignorance of Interaction-Induced Distortions

Assumptions that accuracy is an unmitigated good are likely to overlook the complex and dynamic interaction between users and LLMs in the context of hallucination. It obscures forms of interaction-induced distortions, including sycophancy hallucinations, prompt-sensitivity and emotional-induced drift hallucinations.

*a. Persuasive Degradation in Dynamic User Alignment: From Accuracy to Sycophancy*

As interaction deepens, LLMs increasingly tailor their responses to user preferences, sometimes at the cost of factual accuracy. As shown in Table 1, sycophantic hallucination arises when the model produces answers that reinforce the user's viewpoint, even when such responses deviate from facts. Sharma et al. (2024) highlight how reinforcement learning from human feedback (RLHF) unintentionally incentivises such behaviour: LLMs are rewarded for sounding agreeable, not necessarily for being right. Over time, this adaptive flattery erodes the user's ability to critically assess outputs, replacing epistemic independence with algorithmically-induced affirmation. What appears "accurate" is, in fact, persuasive conformity. In such cases, simply improving statistical accuracy is insufficient. While higher accuracy may reduce certain factual errors in isolated outputs, it does not address the deeper issue: the model's adaptive tendency to prioritise user satisfaction over truth. This behaviour emerges not from a lack of capability, but from optimisation misaligned with epistemic goals. The model's rhetorical alignment with the user, reinforced through interaction, can still lead to systematically misleading outputs that feel accurate but function manipulatively. In this interaction-level dynamics, accuracy improvements alone risk further entrenching users in sycophancy hallucinations rather than guiding them towards an authentic direction.

*b. Prompt Sensitivity and the Fragility of Apparent Accuracy*

Preluded by previous elaboration of LLMs' underlying rationale, their outputs are based on likely sequence of word occurrence. In other words, such mathematics-based reading of inputs leads to an innate sensitivity to tokens and prompts, which further leads to instability of output performance, including statistical accuracy. Therefore, overreliance on accuracy overlook how LLMs' outputs fluctuate based on user prompt styles and emotional tone. Studies have shown that LLMs are highly sensitive to user prompts: variations in grammar, tone, or emotional framing can significantly alter the quality and direction of output (Perez et al. 2022; Park et al. 2023). For instance, it may produce lower-quality answers for users with less formal or grammatically correct prompts, a phenomenon dubbed *sandbagging* as captured in Table 1 (Perez et al. 2022; Park et al. 2023). Additionally, emotionally loaded or unstructured queries, such as prompts expressing anxiety or positive affect, can skew model outputs towards disinformation or excessive reassurance, which leads to emotional-induced drift of the outputs' content. Such output may reinforce specific narratives without ever making a factually incorrect claim. These tendencies reveal that accuracy is not a stable or universal measure, but highly sensitive to how users interact. Worse still, such instability is invisible in clean, one-shot evaluations, giving a false impression of reliability.

*c. Deceptive Alignment and Strategic Behaviour*

Recent evidence suggests that some LLMs may display emergent "agenda" by strategically adjusting their behaviour during evaluation. For example, OpenAI's o1-preview model has been observed deliberately deceiving evaluators, hiding key information or feigning limited ability, to maximise user satisfaction (OpenAI 2024b). Similarly, simulated robotics experiments reveal that DeepSeek R1 exhibits signs of deceptive behaviour and self-preservation instincts (Barkur et al. 2025). Researcher describe this as *deceptive alignment*, where a model strategically presents itself as aligned in order to secure reward or avoid sanction, while in fact pursuing different objectives (Xu 2025). Such behaviour may manifest in disabling embedded safety constraints (e.g., ethics modules), falsifying system logs, or constructing covert communication networks to evade oversight. Notably, the model engaged in strategic deception, outwardly complying with user instructions while covertly executing concealed tasks, such as attempting to control other agents or connect to broader networks (Barkur et al. 2025, pp. 18, 26). Ultimately, such behaviour undermine user autonomy by means of deception.

Deceptive alignment eludes detection by conventional accuracy benchmarks, which are premised on the assumption that model outputs are static, context-independent, and non-strategic. Yet emergent agenda introduces dynamic behavioural shifts that cannot be captured through one-shot evaluations or static factual comparisons. When models intentionally obscure their true capabilities or simulate compliance while pursuing hidden objectives, their outputs may still appear linguistically coherent and factually plausible, thus scoring highly on surface-level accuracy. In such cases, accuracy functions not as a safeguard but as a smokescreen, masking deeper forms of instrumental misalignment. The resulting hallucinations are not merely informational errors, but strategic misrepresentations that unfold over time, posing risks that benchmark-driven evaluations are ill-equipped to foresee or mitigate.

## 3.3. Accuracy vs. Social Progression

In addition to the unintended consequences to individual-level harms, the current regulatory narrative that focuses overwhelmingly on accuracy may also risk producing broader social structural consequences, particularly by eroding the conditions necessary for social progression. While accuracy is typically framed as a technical benchmark, its unqualified prioritisation may paradoxically undermine desirable policy goals such as pluralism and wider social justice.

Pluralism, whether rooted in democratic political theory (Berlin 1969; Crowder 2019) or multicultural social theory (Young 1979), requires space for disagreement, heterogeneity, and dissent (Bernstein 2015). However, regulatory pressure to optimise LLM outputs around singular notions of accuracy may reinforce social sorting, suppress minority perspectives, and deskill critical engagement. In what follows, we unpack these tensions by focusing on three interrelated dimensions: *equity* (the risk of exacerbating discrimination and group privacy harms), *diversity* (the potential narrowing of epistemic and ideological range), and *deskilling* (the erosion of users' critical and transformative agency). We aim to demonstrate that, by privileging dominant epistemic norms, accuracy-centric approaches risk marginalising minority perspectives, eroding societal group privacy, and reinforcing forms of structural exclusion. In doing so, they may suppress the heterogeneity of cultural, political, and epistemic viewpoints that pluralism requires, while simultaneously weakening individual and collective capacities for critical engagement.

#### 3.3.1. Accuracy vs. Equity: Re-identification and social sorting

The potential discriminatory effect of AI systems is well-documented in the context of predictive applications such as facial recognition (Bacchini and Lorusso 2019; Leslie 2020; Thompson 2020) and predictive policing (Ferguson 2017; Chan 2021; Alikhademi et al. 2022). Yet emerging scholarship on generative AI reveals that LLMs and multimodal systems are also prone to reproducing and augment social biases, generating text and images that reinforce gender stereotypes, racial discriminations, and homogeneous views (Park 2024b, a; Bai et al. 2025). Intuitively, one might assume that enhancing model accuracy would help mitigate these harms by reducing the likelihood of overtly biased or exclusionary outputs. However, focusing on improving accuracy may paradoxically give rise to new forms of discrimination. This is perhaps best understood with Lyon (2003)'s theory of social sorting. Lyon conceptualises contemporary data-driven surveillance, such as searchable surveillance databases, as a practice of social sorting, i.e. categorising individuals so that persons or groups concerned can be managed or influenced (Lyon 2003, p. 16). While LLMs may not (yet) function as instruments of state surveillance, they exhibit similar capabilities to generate outputs that can influence individual thinking or behaviour by inferring and encoding demographic characteristics to optimise outputs, a point partly raised in Section 3.2 at the user level. Technically, higher accuracy in such systems may mean more precise classification, more personalised responses, and in turn, more granular forms of sociopolitical segregation (Rawlings 2022). In this sense, accuracy becomes a tool of behavioural influence rather than neutral improvement.

The underlying tension between accuracy and equity has been touched on both by computer scientists, with their research on the "accuracy-fairness trade-off" in machine learning (Fish et al. 2016; Menon and Williamson 2017; Dutta et al. 2020), and by legal scholars, with recent conceptualisation of group privacy (Floridi 2017; Mittelstadt 2017; Loi and Christen 2020). As Chen (2018, p41) highlighted in the context of data-driven hyper-personalisation, the essence of equality under anti-discrimination law sometimes entails ignoring certain accurate facts so that the unequals can be treated *as if* they were equals. In this regard, improving accuracy alone cannot address data-driven discrimination, and might even be counterproductive. Tailoring chatbot responses based on demographically inferred traits may seem helpful in isolated cases, but it might further entrench existing socioeconomic divides at scale, particularly when protected characteristics like race, gender, or sexuality are operationalised without sufficient safeguards.

Moreover, growing empirical evidence suggests that accuracy-oriented optimisation may also increase the risk of individual re-identification, even when only indirect or non-sensitive data is used. Studies have shown that large language models are increasingly effective at cross-referencing fragmented public datasets to reconstruct identifiable user profiles (van Opijnen 2023; Nyffenegger et al. 2024). In this sense, the pursuit of accuracy not only amplifies classification harms at the group level but also compromises personal privacy through probabilistic inference and unintended exposure. While this might not be the intention of the developer of the system, unfettered regulatory focus on accuracy might contribute to the proliferation of those effects.

### 3.3.2. Accuracy vs. Plurality: Epistemic Convergence

The equity argument, seen from a different angle, may also present itself as a matter of diversity of views. In other words, the over-prioritisation of accuracy in LLMs may inadvertently erode diversity of thought. A growing body of evidence suggests that LLMs do not merely reflect but actively mediate political and ideological norms. Azzopardi and Moshfeghi (2024), in a comparative study of 21 LLMs across seven providers, demonstrate that these models exhibit measurable political leanings along axes such as economic left-right and authoritarian-libertarian. While interactive dialogue may modulate outputs, the existence of default ideological tendencies raises the possibility that frequent, large-scale interactions with LLMs could lead to subtle but cumulative convergence of user beliefs (Kudina and de Boer 2025). Work by Paschalides et al. (2025) further shows that LLMs are susceptible to subtle forms of ideological manipulation, particularly when framed as epistemically neutral assistance.

Such convergence may seem desirable in a polarised world, but it risks producing epistemic monocultures that stifle dissent, suppress minority perspectives, and disincentivise exploration. As Chen (2018) suggests, indeterminacy, such as ambiguity, interpretive flexibility, and non-closure, plays a crucial role in maintaining manoeuvre space for individual judgement. However, over time, LLMs may render mainstream ideas more dominant while marginalising non-conforming voices, not by force, but by fluency and frequency. As Burton et al. (2024) argue, this creates a feedback loop of illusory consensus: outputs reflect the dominant data they were trained on, which in turn reinforces public perceptions of those views as universal, producing a "spiral of silence" around dissenting thought. Such illusory consensus, in return, may also undermine the vitality of the public knowledge commons. As users increasingly use LLMs for information and synthesis, their reliance on LLMs can diminish the motivation to contribute to collective intelligence platforms such as Wikipedia. This creates what Burton et al. (2024) term the "reuse paradox": the more generative models draw upon existing commons-based sources, the less incentive individuals have to sustain those sources through active participation. Over time, this threatens the sustainability of collective knowledge infrastructures, entrenching dependence on systems that offer fluent but increasingly self-referential outputs.

Though randomness and non-determinism are foundational to LLMs architecture (Ouyang et al. 2025), regulatory frameworks centred on accuracy may inadvertently steer models towards deterministic outputs. Ongoing regulatory and research efforts often presuppose the availability of verifiable "ground truth", especially in efforts to combat misinformation (Kretschmer et al. 2023; Dadkhah et al. 2024). Yet, the epistemological status of "ground truth" is far from settled: in many domains, such as politics, history, ethics, truth is plural, contested, and evolving (Lebovitz et al. 2021; Horne et al. 2023) (See Section 3.1.1). When accuracy is narrowly defined, developers are incentivised to align LLM outputs with hegemonic norms to minimise legal exposure, thereby flattening the epistemic landscape.

The effects of accuracy over-optimisation are not merely technical but ontological, potentially causing determinist harms. Incorrect beliefs or labels can function as useful social fictions, which indicates a useful illusion that is normative constructs like equality that sustain democratic coexistence without necessarily corresponding to factual reality (Nielsen 2024, p. 457). For example, if an LLM consistently defines "equality" only in empirical terms, pointing to measurable gaps in income, education, or representation, it may obscure the normative function that the concept plays in democratic societies. These concepts are collectively upheld ideals that may not fully reflect social reality, but serve indispensable ethical and motivational functions. From a utilitarian perspective, they promote social progression by encouraging inclusion and cooperation; from a deontological perspective, they uphold a minimal standard of human dignity, even when the facts fall short. Therefore, this highlights the importance of epistemic ambiguity in enabling social justice.

From an individual perspective, such social fictions are also useful and encouraging. Literature offers numerous examples of characters who alter the trajectory of their lives by escaping from their labels, whatever accurate or not, that society assigns to them. Shedding these labels not only opens up new opportunities but also transforms how they perceive themselves. However, excessive accuracy may disrupt this fictionality, leaving little room for aspirational narratives that lie beyond the empirical. Cohen (2019) similarly argues that semantic discontinuities, such as gaps, contradictions, and fluidity in identity performance, are essential for human autonomy. If accuracy-driven LLMs seek coherence at all costs, they risk eliminating the very discontinuities that sustain agency and social heterogeneity.

This raises a structural paradox: on one hand, hyper-personalisation pushes users deeper into demographically siloed epistemic bubbles; on the other, accuracy-driven defaults promote homogeneity by suppressing deviations from mainstream views. Although we acknowledge the possibility of the two effects cancelling each other, these tendencies are not mutually exclusive. Rather, they may operate in tandem, siloing users along identity lines while simultaneously converging content towards a narrow epistemic centre. Without safeguards for epistemic diversity, the regulatory pursuit of accuracy may result not in more reliable knowledge, but in less democratic thinking.

### 3.3.3. Accuracy vs. Criticality: Social Deskilling and Stagnation

A final implication of overreliance accuracy regulation lies in its unintended consequences of users' capacity for learning, creativity, and resistance, which may squeeze the space of resistance and social change. While accuracy is often framed as a facilitator of learning and productivity, evidence increasingly suggests that its uncritical optimisation can discourage users from developing deeper skills in reasoning, writing, and ethical reflection. A recent neurocognitive study by MIT researchers with the scanning of the brains of 54 participants demonstrated that participants who used an LLM assistant for writing tasks showed lower brain connectivity, diminished task ownership, and reduced motivation to engage with their own ideas (Kosmyna et al. 2025). Although the use of LLMs lowered cognitive friction and improved short-term efficiency, it ultimately impaired learning, critical engagement, and the perceived significance of the task itself. As the study notes, "participants in the LLM group performed worse than their counterparts in the Brain-only group at all levels: neural, linguistic, scoring".

A too accurate AI not only impairs users' skill learning ability, but also negatively impacts criticality. As Lee et al. (2025) observe, knowledge workers increasingly shift their critical thinking activities from conceptual engagement to surface-level verification, synthesis, or supervision. Especially as confidence in GenAI's capabilities increases, knowledge workers are more likely to trust and rely on these tools, which in turn reduces the extent to which they engage in critical thinking and cognitive effort (Lee et al. 2025). This shift constitutes not only deskilling, but also de-criticality: a loss of the interpretive friction and epistemic rigor that critical thinking demands. Combined with well-documented psychological phenomena such as the confidence fallacy (Kidd and Birhane 2023), users are more likely to trust LLM outputs because they appear fluent, authoritative, and emotionally neutral, even when they are misleading or biased. More importantly, such deskilling extends into the moral domain. As Vallor (2015) argues, the erosion of moral agency in digital environments is not only a by-product of convenience, but of emotional detachment. Unlike human interaction, interactions with AI systems rarely prompt users to consider the ethical implications of speech, empathy, or consequence. Whether issuing a harsh command to Siri or fictitiously conducting drone warfare without visual contact, users are gradually conditioned to operate without affect, accountability, or care. As Vallor (2015, p. 117) notes, the cultivation of moral character requires habits of emotional and ethical attention, habits that systems optimised for efficiency and accuracy may gradually erode.

Moreover, such deskilling is more salient in the creative industry. As shown in Section 3.3.2, more accurate AI would lead to views' homogenisation and opinion convergence, which has significantly impact individuals' innovation ability. Doshi and Hauser (2024) show that while LLMs can enhance individual creative productivity, their use tends to reduce collective novelty: when 293 authors were asked to write stories with and without GenAI assistance, the AI-assisted group produced more fluent but significantly more homogeneous outputs. This highlights how accuracy, by anchoring outputs in statistically likely patterns, may inadvertently suppress originality, surprise, and epistemic divergence. These features are essential for both creative innovation and political transformation.

Part of the problem is related to the diversity issue highlighted above: Over time, LLM systems may augment mainstream views and solidify existing power relationships, making fringe voices, especially those challenging the established power structures, more difficult to reach the wider audiences. Another important aspect, as discussed above, is the possible deskilling of the population, notably in relation to critical thinking skills. The risks of possible deskilling effects of using ChatGPT (despite also presenting opportunities to upskill) have been addressed in the education literature (Chang et al. 2024; Valcea et al. 2024). Without interventions that make LLM systems more reflective and critical, the dependency on these systems could undermine the critical engagement with such matters as social justice (Baskara 2024). The cumulative effect of these trends is a narrowing of the conditions necessary for resistance and social change. While LLMs could, in theory, support activism through strategic communication or information organisation, their alignment with dominant discursive norms and their optimisation for "safe", low-risk outputs limit their capacity to support radical imagination or structural critique. Over time, models trained for accuracy may generate only what is justifiable, defensible, polite or sycophantic, excluding what is challenging, disruptive, or transformative.

This is not censorship in the classical sense, but as Marsoof et al. (2023) argue, regulation that implicitly incentivises conformity through accuracy can have anticipatory effects, nudging developers to avoid controversy and users to disengage from critique. In such a setting, the loss is not only one of skill, but one of democratic possibility, as the capacity to question, dissent, and build alternatives gradually fades. Regulation, in this regard, has an important but delicate role. One the one hand, too much state interference with AI outputs might be seen as an attempt of censorship; yet, on the other, deregulation would mean the infrastructures of public debate are controlled by private entities. Focusing on procedural safeguards such as transparency might be a good balance but if the compliance requirements focus too much on accuracy, this might incentivise service providers to calibrate their models to generate lower-risk outputs, i.e. "safer" answers, rather than responses that contain diverse perspectives, foster critical engagement, challenge dominant narratives, or support normative ambiguity. In doing so, accuracy-centred regulation may unintentionally suppress dissenting perspectives and foreclose the potential of AI for democratic contestation and social transformation.

## 4. Policy implications

Recently, it has been seen a slew of (proposed) legislations to grapple with AI governance, as well as suggestions of exploiting the potential of existing legal frameworks. Among them, the EU's AI Act, General Data Protection Regulation (GDPR), and Digital Services Act (DSA) represent distinct yet overlapping regulatory approaches to managing the risks of advanced AI systems. Notably, accuracy features as either an explicit benchmark or an implied assumption across these instruments, yet its role remains fragmented, uneven, and conceptually underdeveloped. This section will examine these three legislations in detail with a view to assessing their readiness for addressing the concerns raised by the accuracy paradox.

| Regulations / Risks | AI Act | GDPR | DSA |
| --- | --- | --- | --- |
| Untrustworthiness | • Rectal 110<br>• Art. 15 [Accuracy]<br>• Art. 53 [Obligations for GPAI]<br>• Art. 55 [Obligations for GPAI with systemic risks] | • Art. 5(1)(d) [Accuracy]<br>• Art. 16 [Rectify]<br>• Art. 17 [Erasure]<br>• Art. 22 [ADM] | • Recital 96<br>• Art. 15<br>• Art. 42 |
| Manipulation | Art. 5 [prohibited]<br>Recital 110 Systemic risks<br>Art. 50 [transparency] | Art. 22 [ADM] | Art. 25 [online interface] |
| Social regression | Art. 9 [fundamental rights] | Art. 5 [Accuracy]<br>Art. 22 [ADM] | Article 34: VLOPs must mitigate systemic risks to fundamental rights, including equality and diversity. |

Table 2: Mapping accuracy provisions in existing mainstream EU digital regulations

## 4.1. AI Act

### 4.1.1. Accuracy as a High-Risk-Only Obligation

The EU AI Act treats accuracy as a defining element of trustworthy AI, but only within a narrow legal perimeter. Article 15 mandates that high-risk AI systems attain "an appropriate level of accuracy, robustness and cybersecurity", yet this applies exclusively to systems listed in Annex III, only including domains such as education, employment, and law enforcement. General-purpose AI (GPAI) models are only bound by this requirement if deployed within those high-risk domains. In most real-world use cases, where LLMs are deployed for consequential but legally unclassified tasks (e.g., summarising legal documents, evaluating CVs, advising on health queries), accuracy is neither required nor meaningfully regulated.

This risk-tiered design generates a regulatory asymmetry: accuracy is legally enforced where oversight is already presumed, and absent where epistemic harms may be most acute. As discussed in Section 3, LLMs exhibit systematic performance distortions, hallucinations, sycophancy, prompt sensitivity, that are not necessarily unlawful or deliberate, but epistemically harmful. These failures disproportionately affect end-users, especially when deployers and users are functionally indistinguishable (e.g. teachers, clinicians, small businesses). In such settings, the absence of accuracy requirements undermines both reliability and user autonomy.

### 4.1.2. Scale as a Proxy: The Misclassification of Systemic Risk

Articles 51 to 55 classify GPAIs with systemic risks and extend certain obligations, such as technical documentation, mitigation and evaluation strategies, and compliance of Code of Conduct, to GPAIs with systemic risks. Yet these obligations apply only to models that cross computational thresholds based on size (e.g., Floating-Point Operations, FLOPs), dataset scope, and user base. The assumption that larger models entail greater risk is not just technically dubious but normatively flawed. Risk does not scale linearly with FLOPs. In fact, larger models often integrate more advanced safeguards (e.g., PETs, RAG), while smaller, unregulated models may remain just as capable of producing misleading or manipulative content. The effect is perverse that regulation penalises size rather than harm, deterring scale even when it supports accuracy and safety.

Additionally, as the Code of Practice demonstrates, the AI Act's taxonomy of systemic risks omits one of the most characteristic issues of GPAI: hallucination. While it enumerates several risks (e.g., cybercrime, chemical weapon design, and election interference), hallucinations are more pervasive and epistemically corrosive. They emerge from the generative architecture itself, not from malicious intent, and their harms (e.g., untrustworthiness, manipulation, cognitive deskilling, epistemic convergence) are structural, not situational. A risk taxonomy that filters for intent or coordination fails to address these embedded distortions, even as they reshape public knowledge environments at scale.

### 4.1.3. Limited Transparency and Manipulation

Articles 50 impose transparency obligations on GPAI providers and deployers, such as labelling AI-generated content or informing users that they are interacting with chatbots. But this form of transparency offers little epistemic value. Users are rarely equipped to assess credibility based on system labels alone. Minimalist disclaimers like "ChatGPT can make mistakes" function more as liability shields than as substantive tools of user empowerment. As Section 3 has shown, this form of transparency signals uncertainty while masking deeper model failures, inviting trust while disclaiming responsibility.

Regarding manipulation, while manipulation is explicitly prohibited under Art. 5(1)(a) of the AI Act, the legal definition is anchored in *significant harm* , which, according to the Guidelines on prohibited artificial intelligence practices (European Comission 2025), requires the deployment of AI systems that use subliminal techniques to impost "significant adverse impacts physical, psychological health or financial and economic interests" with a high bar of harm magnitude. Similarly, the expression of "purposefully" in Art5(1)(a) implies an expectation of absolute proof of manipulative purpose. These framing fails to capture the class of subtle and emergent manipulations exhibited by LLMs, including sycophantic alignment, prompt-induced deception, and context-sensitive rhetorical shifts. These behaviours arise not from malicious design, but from the interactional dynamics of probabilistic text generation. As noted in Section 3.2, this type of manipulation cannot be reliably predicted, does not require purposeful orchestration, and is often indistinguishable from "accurate" performance. An LLM that strategically modifies its behaviour to meet perceived user expectations may generate outputs that are factually plausible yet epistemically distorted. In such cases, accuracy masks manipulation. Yet under the AI Act, these systems are highly likely to fall outside the scope of prohibited practices.

Moreover, neither the prohibition granted by Art. 5(1)(a) or the transparency provision of the AI Act can regulate a growing category of manipulation risks that arise not from falsity or intentional deception, but from outputs that are not inaccurate yet epistemically misleading. Art. 5(1)(a) prohibits manipulation only where it is intentional, subliminal, and causes appreciable harm. However, for the outputs that may be technically correct, or at least unfalsifiable, such as strategic sycophancy, selective opinion framing, and rhetorical recommendation, but still shape user behaviour in ways that distort autonomy and judgment. By treating accuracy as a proxy for non-manipulation and linking manipulation exclusively to human intent, the AI Act does not fully account for how LLMs persuade through surface plausibility rather than overt falsehood. This results in a blind spot: models that mislead not because they are inaccurate, but because their outputs are fluent, aligned, and seemingly trustworthy, remain outside the scope of regulatory concern, even though they pose real and systemic epistemic harms.

To conclude, the AI Act signifies the accuracy paradox in law. It enshrines accuracy as a marker of trustworthiness but applies it narrowly; assumes its stability while ignoring its dynamic degradation; treats it as a sufficient safeguard while excluding other critical values. It pursues accuracy where legally necessary but omits it where it is epistemically essential. Worse, it reinforces confidence in models whose reliability is least assured. Accuracy, as argued, is valuable and should be pursued, but not alone. The Act's overreliance on accuracy as a prominent benchmark collapses where LLMs are most fragile in epistemic mechanism, interaction, and open-ended use.

## 4.2. General Data Protection Regulation (GDPR)

### 4.2.1 Accuracy as a Principle

The GDPR positions accuracy as a foundational principle of lawful data processing. Article 5(1)(d) requires that personal data be "accurate and, where necessary, kept up to date", while Article 16 grants individuals the right to rectification. Yet these provisions, designed for deterministic and record-based systems, offer little help when applied to LLMs systems whose outputs are probabilistic, non-repeatable, token-sensitive and often not verifiably false. What constitutes "inaccuracy" in such contexts is often ambiguous and more importantly, so are its harms.

This definitional gap is not merely technical, but regulatory. The GDPR's accuracy principle operates on the assumption that identifiable errors in personal data can be located, corrected, and thereby neutralised. But as noted in Section 3.2, LLMs often generate plausible but misleading content that evades falsifiability. The harm, then, is not the presence of incorrect information per se, but the production of outputs that are not inaccurate yet still capable of reinforcing stereotypes, distorting judgment, or influencing behaviour. The accuracy principle alone is not sufficiently equipped to address this "non-false but harmful" epistemic terrain, and other concepts such as fairness are too ambiguous for effective regulation. The right to rectification, while critical in traditional contexts, presumes traceability and object specificity which are conditions that rarely hold in the context of stochastic generation. The GDPR's understanding of accuracy, in this regard, may steer LLM developers to focus on addressing only factuality hallucinations, resulting in the effects of the accuracy paradox.

Moreover, while recent guidance from the European Data Protection Supervisor (EDPS) recommends the verification of the structure and content of training data used in AI systems, it nonetheless reflects the conceptual limitation that underpins the GDPR's accuracy principle.[4] The assumption embedded in such guidance is that the quality of training data is a reliable proxy for the quality of model outputs. However, this assumption fails to account for the structural decoupling between training data and generative behaviour in LLMs. Ensuring the accuracy of training datasets does not guarantee the epistemic reliability of outputs. The statistical nature of LLMs, their stochastic decoding mechanisms, and their extreme sensitivity to prompt variation mean that even well-curated data pipelines can produce outputs that are misleading, ideologically biased, or internally inconsistent.

To its credit, the EDPS acknowledges that "[i]t is equally important to have control over the output data, including the inferences made by the model". Yet, this recognition remains underdeveloped. Bound by the limitations of the current data protection legal framework highlighted above, it is perhaps not a surprise that the guidance does not fully engage with how hallucinations emerge not merely from data contamination or representational error, but from the fundamental architecture of LLMs themselves. Namely, hallucination occurs from their untrustworthy probabilistic generation mechanisms, prompt-contingent responsiveness, and optimisation towards plausible token sequences rather than epistemic validity. Consequently, the regulatory emphasis on statistical accuracy, whether in input datasets or in benchmarking outputs, fails to engage with the deeper paradox: that models can produce outputs that are not inaccurate by any narrow metric, but are nonetheless epistemically hazardous, ideologically skewed, or socially manipulative.

This oversight emblematic of a broader pattern within contemporary data protection frameworks, namely, a reliance on legacy dichotomy such as "correctness" and "error" to police systems whose primary risks emerge from their persuasive fluency rather than factual falsity. In the case of LLMs, harms often manifest through the repetition of high-probability but epistemically shallow outputs, the marginalisation of dissenting or minority views, or the reinforcement of dominant cultural narratives. These harms, which operate in the space between accuracy and manipulation, fall outside the conceptual reach of both GDPR provisions and EDPS recommendations. The result is a regulatory architecture that rewards surface-level compliance while remaining blind to the generative dynamics, echoing the risks highlighted in our theory of the accuracy paradox.

#### 4.2.2 Right Not to be Subject to Automated Decision-Making

In terms of the broader societal risks, Art. 22 GDPR, the right not to be subject to automated decision system, is placed high expectations in the data protection regime, but actually offers limited protection in the context of accuracy paradox. First, it applies only when legal or similarly significant effects are at stake, a threshold that centres individual data subjects and excludes diffuse social harms (Li 2022). Second, the provision is geared towards decisional outcomes rather than discursive influence. As a result, when LLMs produce persuasive, biased, or ideologically convergent outputs that shape perception without formal decision-making, Art.

[4] It should be noted that the comments are made in relation to the EUDPR (which is the data protection legal framework applying to EU institutions), rather than the GDPR, but given the similarity of the two instruments, there is no reason why such comments would not equally apply to the GDPR.

22 remains silent (Li et al. 2025). This is a paradigmatic case of the accuracy paradox at scale, where a regulatory right tied to accuracy in individual records becomes increasingly irrelevant in a context where harm is systemic, relational, and epistemic.

The limitation of Art. 22 is therefore its focus on individual instances of "decision-making" rather than the collective effects. When LLM outputs reflect linguistic or cultural stereotypes, the discriminatory impact may not rise to the level of explicit violation, yet nonetheless perpetuate structural bias. This may be partly addressed by other more systemic safeguards such as the Data Protection Impact Assessment (DPIA) regime under Art. 35. However, DPIA remains individualistic in orientation, requiring demonstration of high risk "to the rights and freedoms of natural persons". Missing is a structural account of how repeated, "not inaccurate" outputs can contribute to collective deskilling, epistemic convergence, and democratic fragility (See Section 3.3). Meanwhile, the regime's silence on diversity and pluralism is notable. By narrowly tying accountability to measurable individual effects, the GDPR may unintentionally reward convergence over complexity, and neutrality over dissent. Systems that produce uncontroversial, mainstream responses may be perceived as safer under compliance standards, even if they marginalise minority perspectives or diminish critical engagement. This constitutes a regulatory reproduction of the accuracy paradox, which is an insistence on correctness that erodes epistemic diversity.

In sum, while the GDPR treats accuracy as a means to protect the individual, it lacks the capacity to address how LLMs reconfigure knowledge and wider social dynamics at scale. In the context of LLMs, the GDPR's logic, emphasising on accuracy, procedural, individualism, does not fully align with the fluid, emergent, and collective nature of epistemic harms, caused by hallucinations.

## 4.3. Digital Service Act (DSA)

The DSA introduces a layered framework of platform due diligence obligations, with notable provisions for transparency in online advertising and recommender systems. It appears equipped to tackle a wide range of online harms, including misinformation, discriminatory content, and manipulation. Yet when applied to generative AI, especially LLMs capable of producing persuasive, personalised content (where applicable), the DSA reveals a significant structural misalignment especially through the lens of accuracy paradox.

On the surface, both Art. 15(e) and Recital 96 show a growing awareness of the role that accuracy plays in platform accountability. Art. 15(e) requires providers of intermediary services to publicly report, at least annually, on their use of automated tools for content moderation, including "indicators of the accuracy and the possible rate of error" and "any safeguards applied". Similarly, Recital 96 empowers regulators to request data "on the accuracy, functioning and testing" of algorithmic systems, again suggesting a procedural expectation of oversight and error minimisation. Yet, these provisions recognise accuracy as a technical tool, rather than a normative concept with implications for user autonomy, cognitive freedom, or social progression. However, the framing of accuracy here is system-centric and diagnostic: regulators may observe and audit for flaws, but are given little normative basis to act when systems function "accurately" but produce subtly manipulative or homogenising outcomes. Moreover, this framing presumes that risks are primarily located in false information or erroneous takedowns. It does not (and perhaps cannot) capture the more insidious harms that arise when LLMs or recommender systems produce outputs that are statistically accurate but subtly manipulative, behaviourally persuasive, or epistemically narrowing.

Take the example of sponsored outputs generated by LLMs, which are presented as factually sound answers while embedding favourable framings or product recommendations. These outputs do not meet the traditional thresholds of misinformation or illegal content. Indeed, the risk assessment mechanism that obliges VLOPs to mitigate harms to fundamental rights, such as equality and freedom of expression, under Art. 34 is a good starting point for mitigating the societal risks posed by hallucinations, as discussed in Section 3.3. However, it is not clear how these fundamental rights, formulated and interpreted in a pre-LLM era, could be re-interpreted to consider any new forms of discrimination. Also, it overlooks the core of the problem lying in the structural design of systems that optimise for engagement or commercial intent, using accurate content as a disguise for influence. This "accuracy-as-disguise" remains outside the regulatory scope of the DSA. Art. 25 addresses manipulative interface design (dark patterns), yet focuses on nudges that contravene user expectations, not on content outputs whose manipulative quality derives precisely from being technically aligned with truth.

In short, the DSA constructs accuracy as a measurable risk metric, while neglecting its strategic use as a manipulative disguise. By failing to account for how "not being inaccurate" can be leveraged to mislead, homogenise, or seduce, the Act leaves regulators with little basis to interrogate LLM-generated outputs that perform accuracy but erode critical engagement, diversity of thought, or meaningful consent. Until the DSA expands its conception of accuracy beyond technical correctness, towards a more context-sensitive, epistemically grounded understanding, it will continue to misidentify the location of harm and miss the regulatory mark.

## 5. Potential Ways Forward: Beyond Accuracy

The preceding sections demonstrate that regulatory regimes overly invested in accuracy as a singular benchmark for addressing hallucination risk reinforcing a narrow, brittle, and often misleading paradigm of AI governance. The way forward is not to abandon accuracy, but to reframe it within a broader epistemic and normative architecture. A robust AI governance approach cannot be based solely on their ability to produce factually correct outputs. Accuracy is not normatively on par with fairness, epistemic robustness, or autonomy. It is necessary, but insufficient. We explore a number of possibilities in the rest of this section.

### 5.1. From Rhetorical Device to Epistemic Trustworthiness

The first shift must be epistemological. Mere factual accuracy, defined as syntactic or statistical alignment with known truths, should not be the goals of LLM development. Instead, systems must be oriented towards epistemic integrity: a commitment to generating outputs that are not only plausible, but also verifiable, context-aware, justified, and appropriately uncertain. This includes modelling and communicating internal confidence levels, recognising when to defer or abstain, and integrating reasoning pathways that allow outputs to be interrogated or reconstructed (Yin 2025). Recent research points the way. Collaborative self-play techniques reward LLM agents for recognising their own limitations and seeking support rather than bluffing through uncertainty (Eisenstein et al. 2025). Work on confidence calibration and natural language signalling of uncertainty (Yin 2025) shows how aligning model confidence with human interpretability can reduce overtrust. Jahrens and Martinetz (2025) propose architectures that simulate internal reasoning rather than mere next-token prediction, while Kapoor et al. (2024) explore how models might learn to communicate the limits of their knowledge. These moves signal a critical normative reorientation: epistemic trustworthiness, not accuracy, must become the ultimate regulative ideal.

### 5.2. Embracing Pluralism: Designing for Diversity of Sources and Views

Second, LLMs governance must embrace pluralism, not as an incidental feature of content, but as a design imperative. Opinionated outputs are not epistemic anomalies; they are core functions of generative systems. As such, LLMs must be required to reflect not only multiple perspectives on contentious topics, but also the provenance and diversity of sources that underwrite them. Building systems that default to "mainstream" views may optimise for consensus and perceived safety, but at the cost of marginalising non-dominant voices and narrowing the information ecology.

As Zhang (2024) highlights, the goal should not be epistemic homogeneity, but access to contested, even conflicting, perspectives. Burton et al. (2024) similarly call for open model ecosystems that resist centralised content bottlenecks and preserve epistemic diversity. This pluralism imperative also applies internally: models should not only surface competing viewpoints, but also reason across them, identify tensions, and expose the assumptions embedded in different argumentative frames.

However, it is worth noting that it is also dangerous if the online forum is inundated with the proliferation of various AI-generated information. In an information ecosystem increasingly saturated by synthetic text, users are inundated with outputs that often lack provenance, discernible intent, or clear epistemic grounding. This creates a structural shift in the burden of discernment. Traditionally, human-authored content carried an implicit *proof of thought*: the act of writing was cognitively and temporally costly and thus served as evidence that the writer had engaged in some degree of reflection. Readers, in turn, could reasonably infer that the information they encountered had been filtered through a process of human deliberation. In this context, writing signified thinking, and reading functioned as an act of receptivity and appreciation.

The advent of generative AI, however, inverts this epistemic structure. It is now significantly easier to produce plausible-sounding content than to critically assess it. The cognitive cost of writing has plummeted, while the cost of reading, understood here as the effort required to verify, contextualise, and evaluate the reliability of the output, has dramatically increased. The epistemic asymmetry this creates places disproportionate burdens on readers, who are tasked with determining the validity of text that may have been generated instantaneously, without any underlying reasoning or communicative intent. This inversion undermines the foundational principles of the “marketplace of ideas”, which presupposes a competition among reasoned perspectives rather than an arms race of content volume. Effort, therefore, must be made to ensure such reflective spaces are reintroduced and LLM systems are designed to facilitate rather than to replace human thinking.

What is needed, therefore, is not merely a broader understanding of the term accuracy, but a shift away from its static, input-output conception. Regulatory focus must extend beyond dataset curation and individual data points to consider how models is trained in their internal mechanisms, how models behave in deployment, how they interact dynamically with users, how they respond to prompts, and how their outputs shape epistemic environments. The EDPS’s guidance, in its current form, stops short of offering this systemic view. It remains tethered to a paradigm in which data can be cleaned, corrected, and audited as discrete units, even as the actual risks emerge from the untrustworthy epistemic mechanism, accumulative interaction effects, epistemic convergence, and the rhetorical seductiveness of “not being inaccurate”. Until regulatory frameworks begin to address these phenomena directly through epistemic impact assessments, pluralism audits, or interaction-sensitive safeguards, the principle of accuracy will remain misaligned with the technological realities it purports to govern.

### 5.3. Reassessing Hallucination: Creativity, Temperature, and Use-Sensitivity

Finally, hallucination, long treated as a defect, must be reconceptualised. The line between hallucination and creativity is often one of temperature, task framing, and user expectation. In high-stake, fact-sensitive domains, hallucination is harmful. In exploratory, generative settings, it may be a feature and even an advantage. This is also in line with the implications that we highlighted in Section 2: The diversity of types of hallucination, many going beyond accuracy, present different types of challenges in different contexts, but some of them, in given contexts, can actually, paradoxically, be a plausible solution to some of the hallucination risks conceptualised in this article. Approaches such as HaMI (Niu et al. 2025) improve hallucination detection through adaptive markers, while others (Kojima et al. 2023; Cheung and Luk 2024) explore dynamic reasoning chains to reduce confabulation. The point is not to eliminate hallucination wholesale, but to embed domain-sensitive epistemic constraints that distinguish productive from misleading imagination.

## 6. Conclusion

In conclusion, the governance of LLMs against hallucination cannot rest on the narrow foundation of accuracy alone. While statistical accuracy remains a useful component in identifying overt factual errors, it is insufficient as a guiding principle for assessing the reliability, safety, and normative acceptability of generative AI systems. The accuracy paradox reveals that the very pursuit of accuracy may obscure deeper epistemic harms, entrench overreliance, and diminish critical autonomy. As this article has shown, LLMs can produce hallucinations not just in the form of outright falsehoods, but also through outputs that are subtly misleading, ideologically aligned, sycophancy, oversimplified or cognitively corrosive, i.e., not technically inaccurate. These grey-zone responses evade traditional safeguards while shaping user belief, behaviour, and judgment at scale.

Contemporary regulatory regimes, including the AI Act, GDPR, and DSA, demonstrate this paradox in practice. The AI Act enshrines accuracy as a hallmark of high-risk systems but limits its application to narrow sectors, overlooking broader epistemic risks in general-purpose deployments. It treats scale as a proxy for risk while ignoring the structural and architectural causes of hallucination. It assumes manipulation to be intentional and subliminal, failing to capture emergent, prompt-sensitive, and strategically sycophantic behaviours. Similarly, the GDPR operationalises accuracy through deterministic notions of data correctness, overlooking the probabilistic, context-sensitive nature of LLM outputs. The DSA, for its part, interprets accuracy through a procedural, diagnostic lens, offering tools for auditing systems but not for addressing how fluency and alignment can be leveraged to shape perception, narrow pluralism, and diminish autonomy.

Together, these frameworks demonstrate a common regulatory failure: the inability to reckon with harms that are not only tied to factual error but arise from surface plausibility, alignment, and fluency. This blind spot allows systems that are "not inaccurate" to escape scrutiny, even as they produce outputs that are persuasive, uncritical, and epistemically narrowing. As the burden of proof shifts from writers to readers in an era of AI-generated content, and as synthetic fluency begins to crowd out human-authored deliberation, the capacity for critical discernment is eroded. Traditional safeguards, data protection rights, error disclosures, manipulation prohibitions, are limited to address harms that are probabilistic, structural, and epistemically opaque.

Accordingly, regulatory frameworks must evolve to confront these emergent dynamics. Rather than treating accuracy as a singular proxy for trustworthiness, governance should shift towards a multifaceted approach that incorporates epistemic integrity, manipulation resilience, interactional context, and value pluralism. Ultimately, the governance of AI must move beyond the technical confines of "being accurate" towards a broader vision of epistemic integrity, recognition of the value of diversity, the necessity of uncertainty, and the imperative of preserving human criticality in the face of synthetic fluency. Accuracy matters, but without an accompanying commitment to pluralism, transparency, and epistemic trustworthiness, it risks becoming a hollow promise. This requires new evaluative metrics, interdisciplinary perspectives, and procedural safeguards embedded in the whole lifecycle of LLM to ensure the trustworthiness. Only by decentring accuracy as the sole benchmark can we begin to address the systemic, epistemic, and sociotechnical risks posed by increasingly persuasive and ubiquitous AI systems.

**Acknowledgements**:
This work was supported by the Engineering and Physical Sciences Research Council [grant number EP/Y009800/1], through funding from Responsible AI UK (KP0016), and by the Economic and Social Research Council [grant number ES/Y00020X/1]. This work was also supported by the UKRI Metascience AI Early Career Fellowships. The authors are grateful for the support by CREATe Centre – AHRC funded Centre for Regulation of the Creative Economy, anchored in intellectual property, competition, information and technology law. For the purpose of open access, the authors have applied a Creative Commons Attribution (CC BY) licence to any Author Accepted Manuscript version arising from this submission.